\newcommand{\CLASSINPUTtoptextmargin}{0.75in}
\newcommand{\CLASSINPUTbottomtextmargin}{1in}
\documentclass[conference]{IEEEtran}

\usepackage{amsmath}
\usepackage{amsfonts}
\usepackage{commath}
\usepackage{enumitem}
\DeclareMathOperator*{\argmax}{arg\,max}

\usepackage{array}

\usepackage{url}

\usepackage{colortbl}
\usepackage{xcolor}
\usepackage{tabularx}
\usepackage{graphicx}
\usepackage{float}
\usepackage{subfig}
\def\BibTeX{{\rm B\kern-.05em{\sc i\kern-.025em b}\kern-.08em
    T\kern-.1667em\lower.7ex\hbox{E}\kern-.125emX}}
\begin{document}
\title{Automatic Response Assessment in Regions of Language Cortex in Epilepsy Patients Using ECoG-based  Functional Mapping and Machine Learning}

\author{\IEEEauthorblockN{Harish RaviPrakash}
\IEEEauthorblockA{Center for Research in Computer Vision\\
College of Engineering and Computer Science\\
University of Central Florida\\
Orlando, Florida 32826}
 \and
 \IEEEauthorblockN{Milena Korostenskaja}
 \IEEEauthorblockA{Functional Brain Mapping \\and Brain Computer Interface Lab\\
 Florida Hospital for Children\\
 Orlando, Florida - 32803}
 \and
 \IEEEauthorblockN{Ki Lee}
 \IEEEauthorblockA{Functional Brain Mapping \\and Brain Computer Interface Lab\\
 Florida Hospital for Children\\
 Orlando, Florida - 32803}
 \and
 \IEEEauthorblockN{James Baumgartner}
 \IEEEauthorblockA{Functional Brain Mapping \\and Brain Computer Interface Lab\\
 Florida Hospital for Children\\
 Orlando, Florida - 32803}
 \and
 \IEEEauthorblockN{Eduardo Castillo}
 \IEEEauthorblockA{MEG Lab\\
 Florida Hospital for Children\\ 
 Orlando, Florida 32803}
 \and 
 \IEEEauthorblockN{Ulas Bagci}
 \IEEEauthorblockA{Center for Research in Computer Vision\\
 College of Engineering and Computer Science\\
 University of Central Florida\\
 Orlando, Florida 32826}}


%

\maketitle

\begin{abstract}
Accurate localization of brain regions responsible for language and cognitive functions in Epilepsy patients should be carefully determined prior to surgery. Electrocorticography (ECoG)-based Real Time Functional Mapping (RTFM) has been shown to be a safer alternative to the electrical cortical stimulation mapping (ESM), which is currently the clinical/gold standard. Conventional methods for analyzing RTFM signals are based on statistical comparison of signal power at certain frequency bands. Compared to gold standard (ESM), they have limited accuracies when assessing channel responses.

In this study, we address the accuracy limitation of the current RTFM signal estimation methods by analyzing the full frequency spectrum of the signal and replacing signal power estimation methods with machine learning algorithms, specifically random forest (RF), as a proof of concept. We train RF with power spectral density of the time-series RTFM signal in supervised learning framework where ground truth labels are obtained from the ESM. Results obtained from RTFM of six adult patients in a strictly controlled experimental setup reveal the state of the art detection accuracy of $\approx 78\%$ for the language comprehension task, an improvement of $23\%$ over the conventional RTFM estimation method. To the best of our knowledge, this is the first study exploring the use of machine learning approaches for determining RTFM signal characteristics, and using the whole-frequency band for better region localization. Our results demonstrate the feasibility of machine learning based RTFM signal analysis method over the full spectrum to be a clinical routine in the near future. 

\end{abstract}


%
\IEEEpeerreviewmaketitle

\section{Introduction}
\begin{figure}[htbp]
\centering
	\includegraphics[width=\columnwidth,height = 5.6cm]{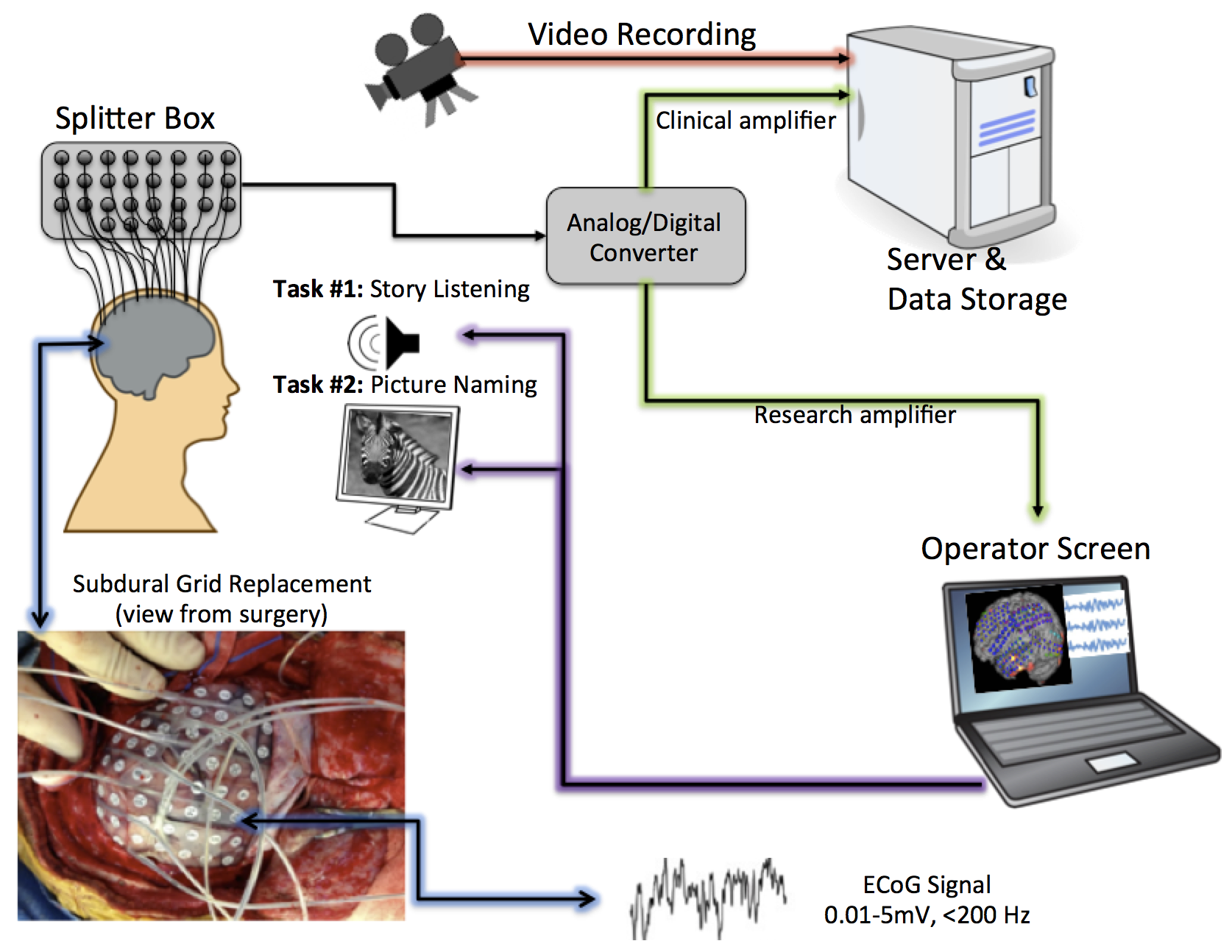}
\caption{The language localization framework with RTFM approach include the following steps: ECoG signal recording, data transfer, storage, research and clinical paths, and tasks. Note that, RTFM signals are obtained from subdurally implanted grid electrodes. \label{RTFM}}
\end{figure}
Epilepsy is a neurological disorder characterized by unpredictable seizures. There are over 65 million people around the world who have epilepsy and an incidence rate of 150,000 new cases every year in just USA alone \cite{epilepsyStats}. Drug Resistant Epilepsy (DRE) (or intractable epilepsy) is defined when the seizures cannot be controlled by medications and about 25\% of all epileptic cases are DRE \cite{shorvon2013longitudinal}. The only viable option in this case is to surgically remove the affected tissue. Epilepsy surgery is a curative option for pharmacoresistant epilepsy, but brain regions associated with language and cognitive functions can be affected by surgery. To do this accurately, unaffected regions of the brain must be identified (called "localized"). The motor and language comprehension are examples of functionally significant region localization. Accurate localization helps to prevent post-surgical loss of functionality.

\subsubsection*{Clinical standard and the state-of-the-art method for RTFM evaluation}
The gold standard task localization, the Electro-Cortical Stimulation Mapping (ESM), utilizes electrodes that are placed on the surface of the brain by means of craniotomy. During the ESM, the current is delivered for a short duration to stimulate the region of interest. The behavioral response corresponding to changes in function are simultaneously recorded. The inherent drawback of this approach is that \textbf{the stimulation can cause the neurons in that region to uncontrollably discharge, i.e., cause seizure.} Recently, ElectroCorticography (ECoG)-based real-time functional mapping (RTFM) \cite{schalk2008real} has been proposed as a promising alternative to ESM. The typical RTFM based task localization and experimental setup is illustrated in Figure~\ref{RTFM}. Similar to the ESM, subdural grids on the cortical surface are utilized for signal collection, however, no external stimulus is provided and only the physiological changes corresponding to the processed stimuli are recorded via the electrodes. Hence, no seizure due to stimulation occurs. 
\begin{figure}
\centering
	\includegraphics[width=\columnwidth,height = 5.3cm]{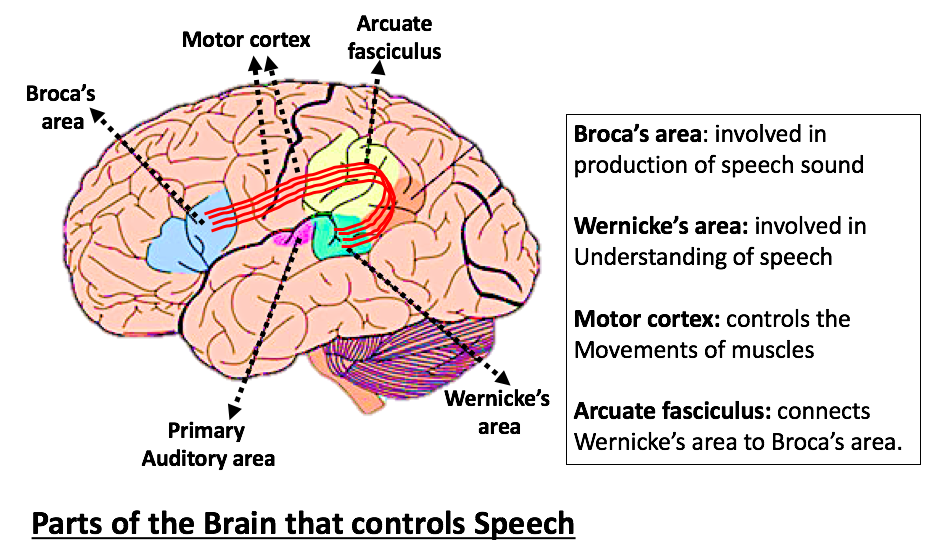}
\caption{Language specific areas in the brain.}
\label{speechcontrol}
\end{figure}
\subsubsection*{Research gap} 
The results of RTFM are not always concordant with the gold standard due to the difficulty in understanding the brain signals without stimulation and lack of sufficient accuracy of the state-of-the-art method, ECoG-based functional mapping~\cite{korostenskaja2015electrocorticography} (ECoG-EM from now on, where EM stands for expectation maximization). There is a need for a method that would improve RTFM signal classification accuracy and make it a strong and safer alternative to the ESM. Current approaches for detecting positive response channels in the eloquent cortex localization task, focus on the power of the signal in the $\alpha$, $\beta$ and primarily, the high-$\gamma$ (70Hz-170Hz) frequency bands \cite{schalk2008real,prueckl2013cortiq}. In these approaches, a baseline recording of each channel at resting-state is used. The power of the signal during the tests is computed using an autoregressive (AR) spectral estimation approach and is then statistically compared to the baseline to calculate the probability whether the channel has a response that is significantly different from it's resting-state (baseline) condition or not. This is repeated every 100 ms for the entire experiment. These approaches do not compare the channels to each other and also do not account for the signal in the frequency range beyond high-$\gamma$.

\subsubsection*{Our contributions}
We present \underline{a novel framework} for ECoG signal analysis with RF to accurately discriminate channels that respond positive and negative in regards to language functional mapping task. To the best of our knowledge, this is the first work comparing the different (positive and negative) responses rather than using a baseline approach. We show the superiority of our approach to the state of the art ECoG-based functional analysis using Expectation Maximization approaches (ECoG-EM), and demonstrate its strong potential to become an alternative to ESM. The rest of the paper is organized as follows: In Sec. \ref{methods} we discuss the ECoG data collection, pre-processing of the data into the discriminative domain and the proposed classification approach. In Sec. \ref{expt}, we present our experimental results. In Sec. \ref{conclusion}, we summarize our findings.

\section{Methods\label{methods}}
\subsection{Data Collection and Experimental Setup}
ECoG represents the electrical activity of the brain recorded directly from the cortical surface. ECoG-based functional mapping allows identification of brain activity correlated with certain task, e.g., language. The basic setup for ECoG-based functional mapping is shown in Figure \ref{RTFM}. ECoG signals from the implanted subdural grids are split into two streams: one for continuous clinical seizure monitoring and the other for ECoG-based functional mapping. The tool used to record the incoming ECoG signal was BCI2000 \cite{bci2000}. A baseline recording of the cortical activity was first acquired to capture the "resting-state" neuronal activity of the regions. 
\begin{figure}[b]
\centering
	\includegraphics[width=\columnwidth,height = 5.3cm]{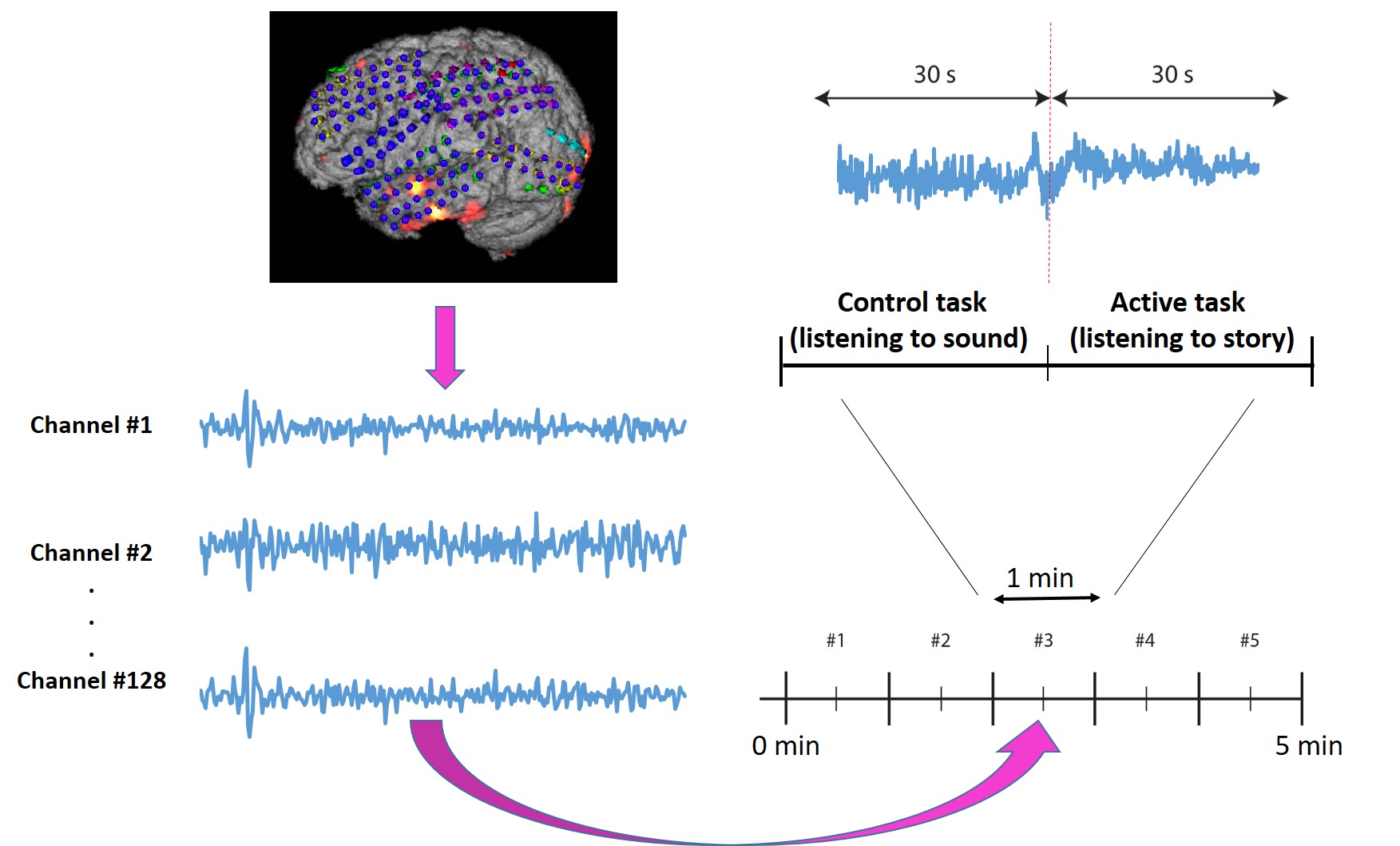}
\caption{Subdural grid localization and position of ECoG electrodes on the brain surface are illustrated (left). For a sample of 1 min duration, both control and active tasks are illustrated (right).}
\label{setup}
\end{figure}
\begin{figure*}[ht]
    \centering
  \subfloat[High Gamma Frequency band - Active task block]{%
       \frame{\includegraphics[width=0.45\textwidth]{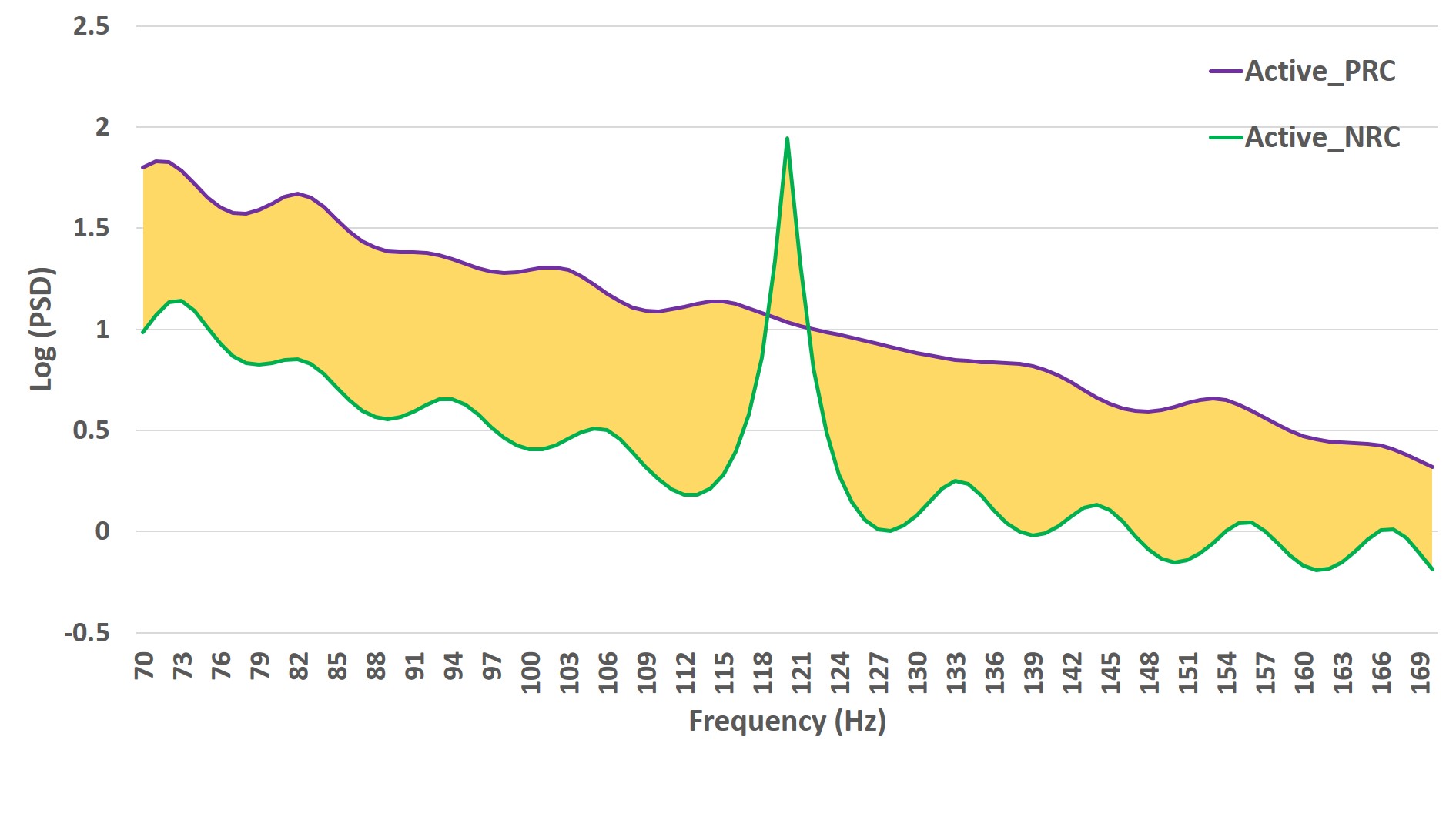}}}
    \label{1a}\hfill
  \subfloat[High Gamma Frequency band - Control block]{%
        \frame{\includegraphics[width=0.45\textwidth]{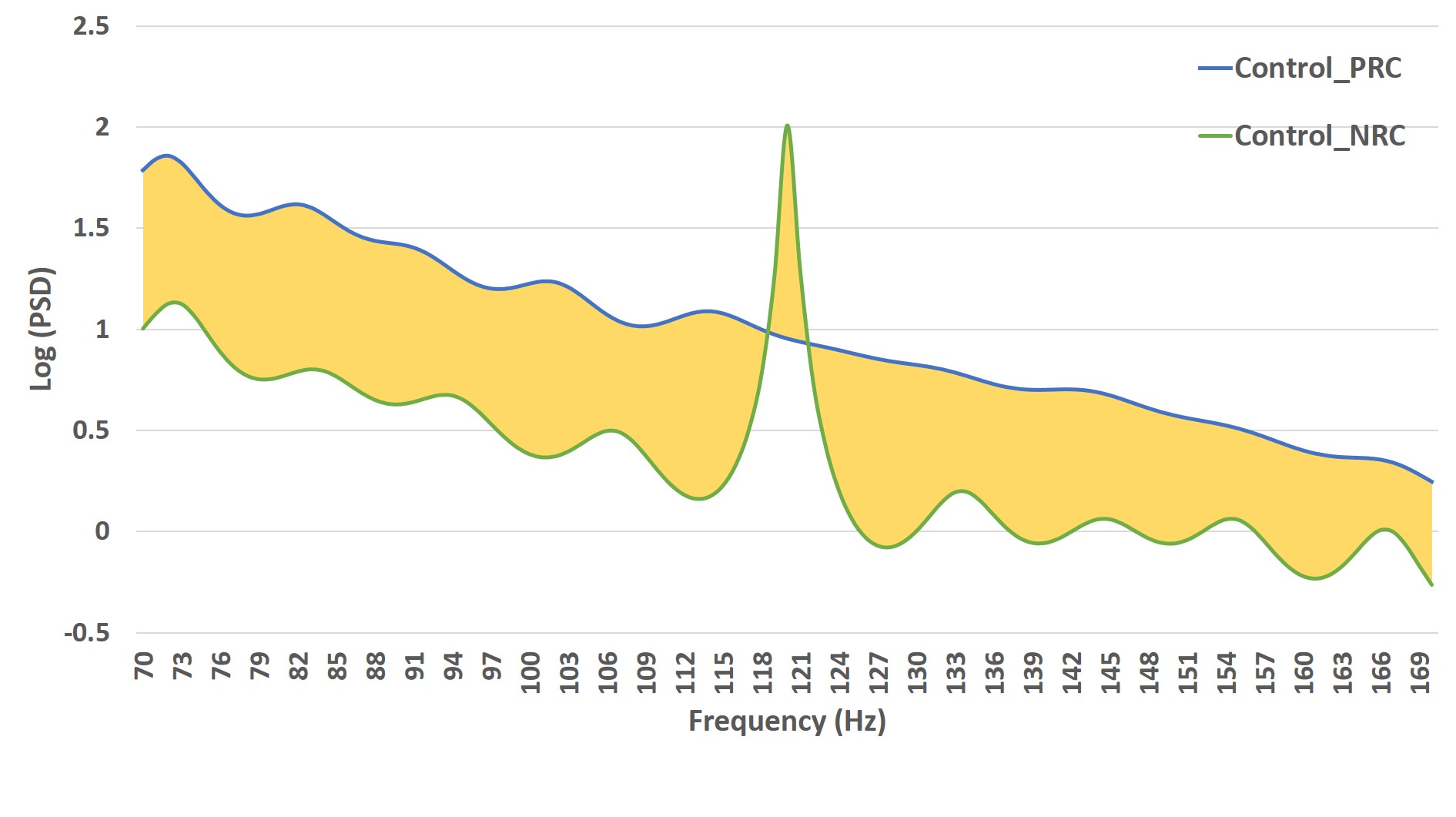}}}
    \label{1b}\\
  \subfloat[Higher Frequency bands - Active task block]{%
        \frame{\includegraphics[width=0.45\textwidth]{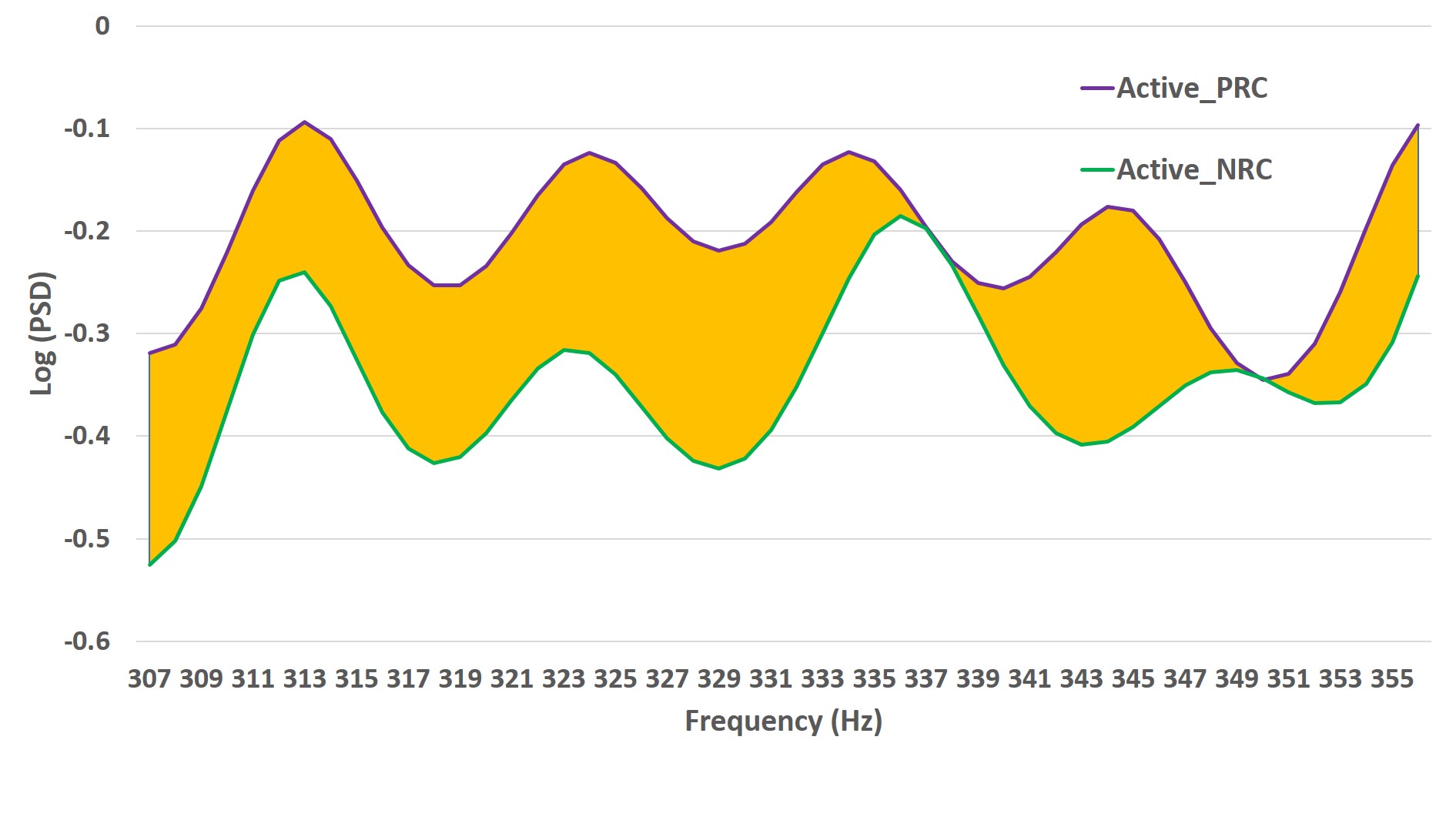}}}
    \label{1c}\hfill
  \subfloat[Higher Frequency bands - Control task block]{%
        \frame{\includegraphics[width=0.45\textwidth]{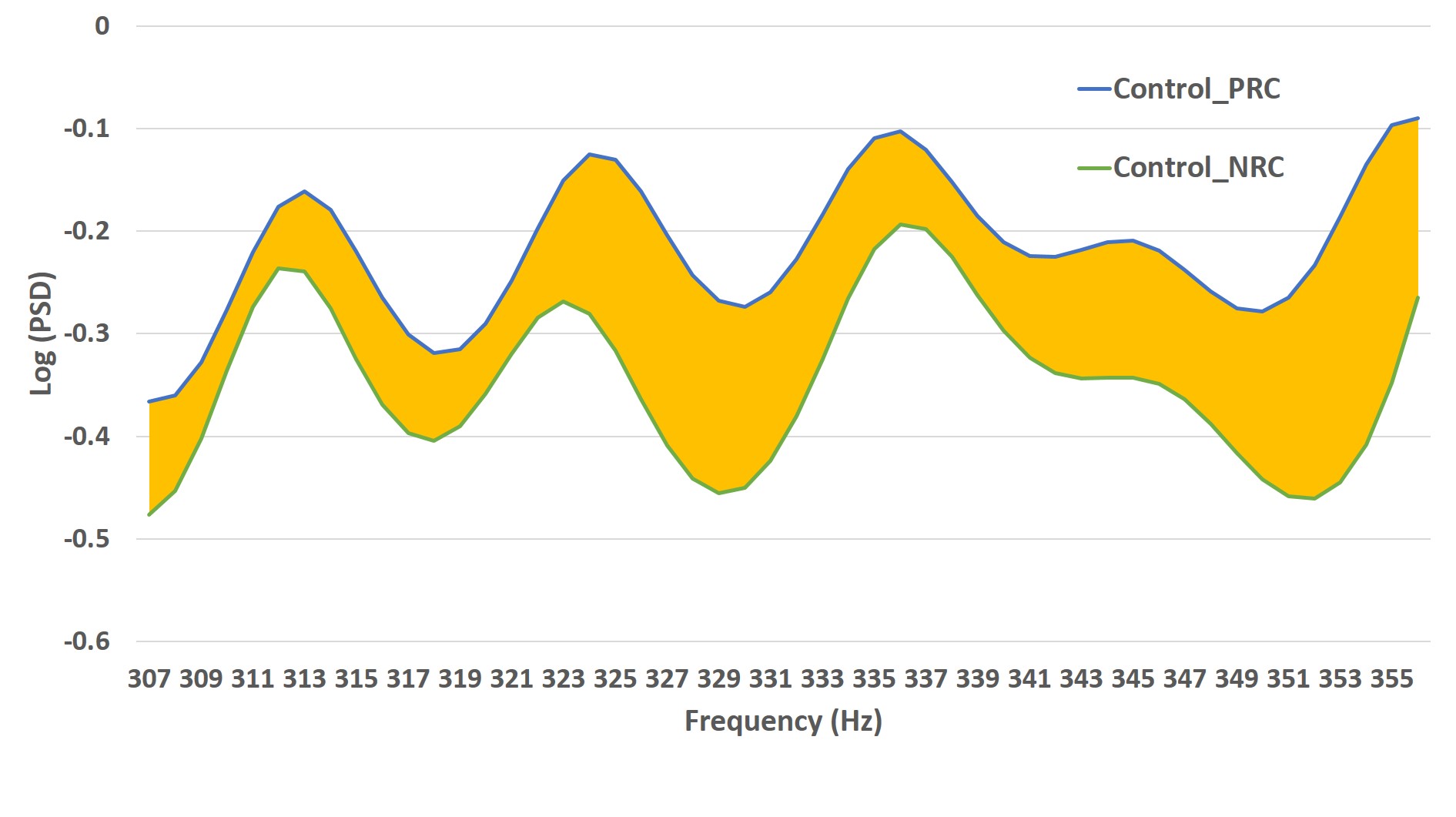}}}
     \label{1d} 
  \caption{PRC vs NRC in different frequency bands. a,b show an example of the difference between PRC and NRC in high-$\gamma$ band. c,d show the same samples in a higher frequency range.}
  \label{freqBands} 
\end{figure*}
The literature on localization of motor function using ECoG-based functional mapping (such as RTFM) is vast \cite{roland2010passive}\cite{kapeller2015cortiq}. Unlike good accuracies obtained from such studies, the localization of eloquent language cortex has proved to be more challenging \cite{arya2015electrocorticographic}. The language function in the brain is processed in several regions primarily, the \textbf{Wernicke's area} and \textbf{Broca's area} as demonstrated in Figure \ref{speechcontrol}.

The Wernicke's area is located in the posterior section of the superior temporal gyrus and is responsible for the receptive language task i.e., language comprehension. The Broca's area, on the other hand, is more involved in speech production. There exists an anatomical connection between these two regions, named the arcuate fasciculus, which could induce a response in one region owing to the other's activation.  
\subsubsection*{Language comprehension task}
Following the baseline recording step, paradigms similar to those employed in ESM or functional Magnetic Resonance Imaging (fMRI) are also employed to record the task-related ECoG signal for functional mapping purposes \cite{korostenskaja}.
Figure \ref{setup} shows one such paradigm, mimicking experimental setup for the language comprehension task. Alternate 30 second blocks of ECoG data during ``control" and ``active" conditions are recorded continuously at a fixed sampling rate of 1200 Hz.
 
For the language comprehension task, the active condition implies listening to a story, while the control task involves listening to broadband noise \cite{korostenskaja2014real}. Another associated paradigm is the reading comprehension task where the subject reads sentences from a screen, and replies with a "True" or "False" response. The system records information from 128 ECoG channels as illustrated in Figure~\ref{setup}.

\subsection{Pre-processing}
As a first step of preparing the data, non-task/control time points in the signal are eliminated. These correspond to the spontaneous activity recording before the 0-min in Figure \ref{setup} and any trailing signals at the end of the experiment. The use of power spectral density (PSD) is proposed in \cite{schalk2008real} as a discriminating feature between the baseline and task signals. In a slightly different manner, we represent PSD with a number of coefficients extracted from an autoregressive (AR) model. Unlike conventional methods, we simplify signal representation with PSD coefficients only. Herein, the AR parameters, $\tilde{a}[n]$, are estimated by forward linear prediction coefficients and then, the spectral estimate is calculated as 
\begin{eqnarray}\label{psd}
\tilde{P}(f) = \frac{T\tilde{\rho}}{\abs{1+\sum_{n=1}^{p}\tilde{a}[n]e^{-i2\pi fnT}}}
\end{eqnarray}
where $T$ is the inverse of the sampling rate ($f_s$), $\tilde{\rho}$ is the estimated noise variance, and $p$ is the order of the AR process. This approach gives us $\frac{f_s}{2}+1$ frequency components. The PSD estimates are computed for each block (task/control) of each channel. Later, we use these components as features to determine RTFM characteristics. 

\subsection{Classification Model}
To differentiate positive response channels (PRC) from negative response channels (NRC), we identify structured signal patterns in signal blocks, which are not readily visible to the human eyes. We hypothesize that the features of the active and control tasks are globally similar between PRC and NRC but still include substantial differences. This hypothesis can be visually tested  and partially confirmed in Figure \ref{freqBands} where the PSD of the active and control blocks of PRC are larger than that of NRC.

To test our hypothesis and provide scientific evidences of ECoG signal separation between functionally positive and negative regions, we design a RF classifier \cite{breiman2001random} to model structured local signal patterns for challenging RTFM signal characterization.  It has been shown in various different areas that RF is an efficient classifier with considerably good accuracies in classification tasks \cite{bromiley2016fully, verhoeven2016using, sarfaraz}. Its superiority to most other classifiers comes from its \textit{generalization} property.
\begin{figure}
\centering
	\includegraphics[width=0.7\columnwidth,height=9cm]{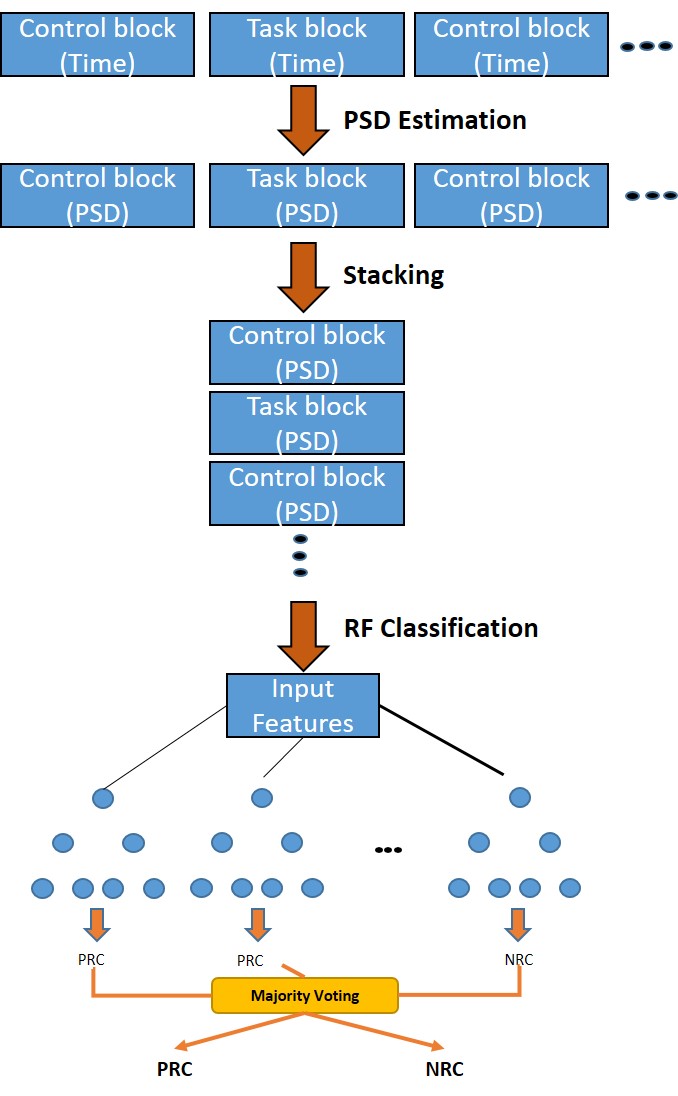}
\caption{Auto-classification workflow\label{rf}: First the signals are split into it's contributing blocks. After, power spectral density (PSD) of the signal is estimated and the blocks are stacked from all channels. Finally, a random forest (RF) classifier is used for discriminating positive response channels (PRC) and negative response channels (NRC).}
\end{figure}
In RF, briefly, each new tree is created and grown by first randomly sub-sampling the data with replacement. An ensemble of algorithms are used so that the sub-trees are learned differently from each other. For a feature vector $\mathbf{v}=(v_1,v_2,...,v_d)\in \mathbb{R}^d$, where $d$ represents feature dimension, RF trains multiple decision trees and the output is determined based on combined predictions. In each node of decision trees, there is a weak learner (or split function) with binary output: $h(\mathbf{v},\theta): \mathbb{R}^d \times \mathcal{T} \rightarrow \{0,1\}$, where $\mathcal{T}$ represents the space of all split parameters. Note that each node is assigned a different split function. RF includes hierarchically organized decision trees, in which data arriving at node $j$ is divided into two parameters. \\
Overall, RF treats finding split parameters $\theta_j$ as an optimization problem $\theta_j=\argmax_{\theta\in\mathcal{T}}{I(\mathbf{v},\theta)}$, where $I$ is the objective function (i.e., split function) and $v$ represents the PSD coefficients in this particular application. As the tree is grown (Figure~\ref{rf}), an information criterion is used to determine the quality of a split. Commonly used metrics are \textbf{Gini impurity} and \textbf{Entropy} for information gain. To overcome potential over-fittings, a random sample of features is input to the trees so that the resulting predictions have minimal correlation with each other (i.e., minimum redundancy is achieved). In our experiments, we have used linear data separation model of the RF.

In our experiments, we use full spectrum of RTFM signal (0-600 Hz) in frequency domain instead of restricted $\gamma$-band. Moreover, we stack the signal to enhance the frequency specific features rather than concatenating them. Each channel has 10 blocks (Figure \ref{setup}) and the final channel classification is based on a majority voting (Figure~\ref{rf}) on the classified sub-blocks. For the tested data point (feature) $\mathbf{v}$, the output is computed as a conditional distribution $p(c|\mathbf{v})$ where $c$ represents the categorical labels (positive vs. negative response). Final decision (classification) is made after using majority voting over $K$ leafs: $p(c|\mathbf{v})=\frac{1}{K}\sum_{k=1}^{K}p_t(c|\mathbf{v})$.

\subsubsection*{Model parameters}
Number of trees, number of features, and data size fed to each tree with or without resampling and the information metric for data splitting are some of the RF parameters that need to be optimized. To achieve this, the model was repeatedly tested under different combinations of the above parameters. For the total number of trees, an incremental update approach was used where we increased the total number of trees till the increase in performance was negligible. Similarly, the number of features was set as the square root of the number of input variables. For the choice of splitting function, Gini impurity was used as for a binary classification problem, both measures yield similar results \cite{breiman1996technical}.

\section{Experiments and Results}\label{expt}
\begin{table*}[!htbp]
\caption{Patient demographics, clinical information, grid placement, and information about the number of analyzed channels are summarized.   Left hemisphere language dominant in all study participants.\label{demographics}}
\centering
\begin{tabular}{|c|c|c|c|c|c|l|}
\hline
\rowcolor{gray!50}
Subject & Age & Sex & Epilepsy & Grid & Epilepsy & Channels Tested/\\ \rowcolor{gray!50}
\# & (yrs) &  &  Focus & Placement & Onset (yrs) & PRC / NRC\\ \hline
1 & 19 & M & Frontal-Temporal & Lateral & 16 & 54 / 22 / 32\\ \hline
2 & 33 & F & Frontal-Temporal & Lateral & 10 & 32 / 5 / 27\\ \hline
3 & 20 & M & Frontal-Temporal & Lateral & 6 & 127 / 16 / 111\\ \hline
4 & 22 & F & Parietal & Lateral & 20 & 30 / 19 / 11\\ \hline
5 & 32 & F & Temporal & Bilateral & 26 & 48 / 10 / 38\\ \hline
6 & 52 & M & Temporal & Lateral & 30 & 48 / 5 / 43\\ \hline
\end{tabular}
\end{table*}
With IRB approval, ECoG data were recorded from six adult patients with intractable epilepsy. Table \ref{demographics} summarizes the patient demographics and the number of channels tested per patient. The ESM results were served as gold standard for separating ECoG channels into two classes: "ESM-positive" and "ESM-negative" electrodes. The number  of tested ESM electrodes varies based on the task in hand (the function that can be compromised during the surgery and therefore needs to be localized), patient's status, possible after-discharges, location of the grid on the brain surface, the epilepsy focus and to a smaller extent on the specialist performing the test. 

Except subject 4, all subjects were tested with the language comprehension paradigm as shown in Figure \ref{setup}. Subject 4, on the other hand, underwent the reading comprehension test involving reading sentences presented on the screen and responding to questions as "True" or "False". Since this test also incorporates speech which would incite a response from face/tongue sensory motor areas of the brain as well as the Broca's area, channels corresponding to these specific regions were not included in our calculations. 

There were $77$ PRCs and $262$ NRCs in total. Each data block in a channel was assigned the same label. For 5 minutes long recording, we had 5 blocks of control and active conditions each per channel and hence, 3390 data samples in total. Due to the large imbalance in data, $77$ NRCs were randomly chosen from the $262$. In total, we have $1540$ blocks of data. For unbiased evaluation of the RF based results, we used 10-fold cross-validation and the average over a 100-iteration was conducted.

\subsubsection*{Time-domain analysis}
First, we tested whether the raw time signal data has sufficiently discriminating information. For this analysis, a RF model with 100 trees was used. The resulting classification accuracy was $61.79\%$ with sensitivity and specificity around $60\%$. While this is marginally better than the simple flip of a coin scenario, it is insufficient to encourage the use of ECoG-based functional mapping over ESM.

\subsubsection*{Frequency-domain analysis}
Each block in a time-domain signal was transformed into the frequency-domain using the pre-processing step described in Section \ref{methods} (i.e., PSD coefficient via AR model). The order of the AR process is set to $SamplingRate/10 = 120$. The PSD estimate is of length $f_s/2+1=$ 601. We then log normalized PSD coefficients to train a RF classifier. An ensemble of 200 bagged classification trees was trained on 9 folds of the data and tested on the last fold.

In order to validate the use of control \& active task blocks for channel classification, we first performed block classification on the $1540$ blocks. The classification accuracy was found to be $94\%$ with sensitivity and specificity of $\approx 93\%$. These results validate the efficacy of the proposed block-based classification strategy.

\subsubsection*{Frequency-band analysis}
Three different experiments (E1, E2, E3) were performed to understand the contribution of the different frequency bands to the channel classification problem:
\begin{enumerate}[label={E\arabic*.}]
\item Classification using full signal spectrum
\item Classification using $\alpha$, $\beta$, high-$\gamma$ sub-bands
\item Classification using only the High-Gamma sub-band
\end{enumerate}
\begin{figure}
\centering
	\includegraphics[width=\columnwidth]{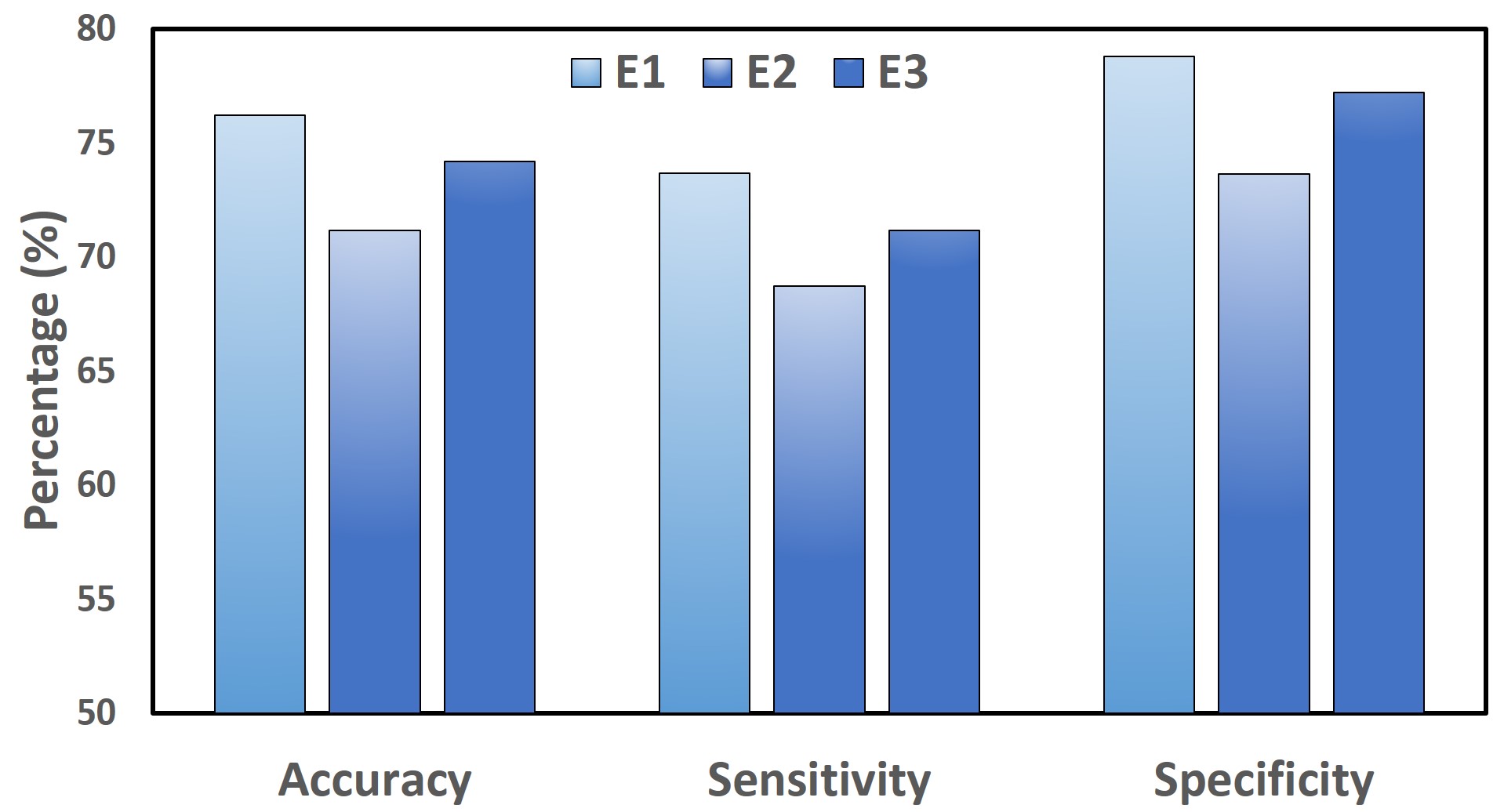}
\caption{Classification scores on ECoG signal classification on Language Comprehension Task.  E1 - Classification using full signal spectrum, E2 - Classification using $\alpha$, $\beta$, high-$\gamma$ sub-bands \& E3 - Classification using only the high-$\gamma$ band. \label{lang}}
\end{figure}
In these experiments, the blocks were classified and a majority voting was applied to classify a channel as PRC/NRC. Figure \ref{lang} summarizes the results of the above experiments for the language comprehension task. In concordance to what was observed in the ECoG-EM approaches such as SIGFRIED \cite{schalk2008real} and CortiQ \cite{cortiq}, we found that the lower frequency bands, specifically, $\alpha$ and $\beta$, did not contribute largely towards classification and the high-$\gamma$ band achieved good classification accuracy. In other words, the full signal spectrum based classification had higher classification accuracy, sensitivity and specificity than the sub-band approaches indicating that the full spectrum had more information to offer. 

\subsubsection*{Block-size analysis}
We also tested the use of smaller blocks of data by further dividing each control/active task block into 10 sub-blocks. Each sub-block of data was the power spectrum representation of 3 seconds of the recording. The classification was done based on a majority voting of the classified sub-blocks within a channel. The resulting classification accuracy was reported to be  $78\%$, higher than the block-based approach. This indicates that there was more local information to be extracted from the signal.

\subsubsection*{Comparison to the state of the art}
\begin{figure}
\centering
	\includegraphics[width=\columnwidth]{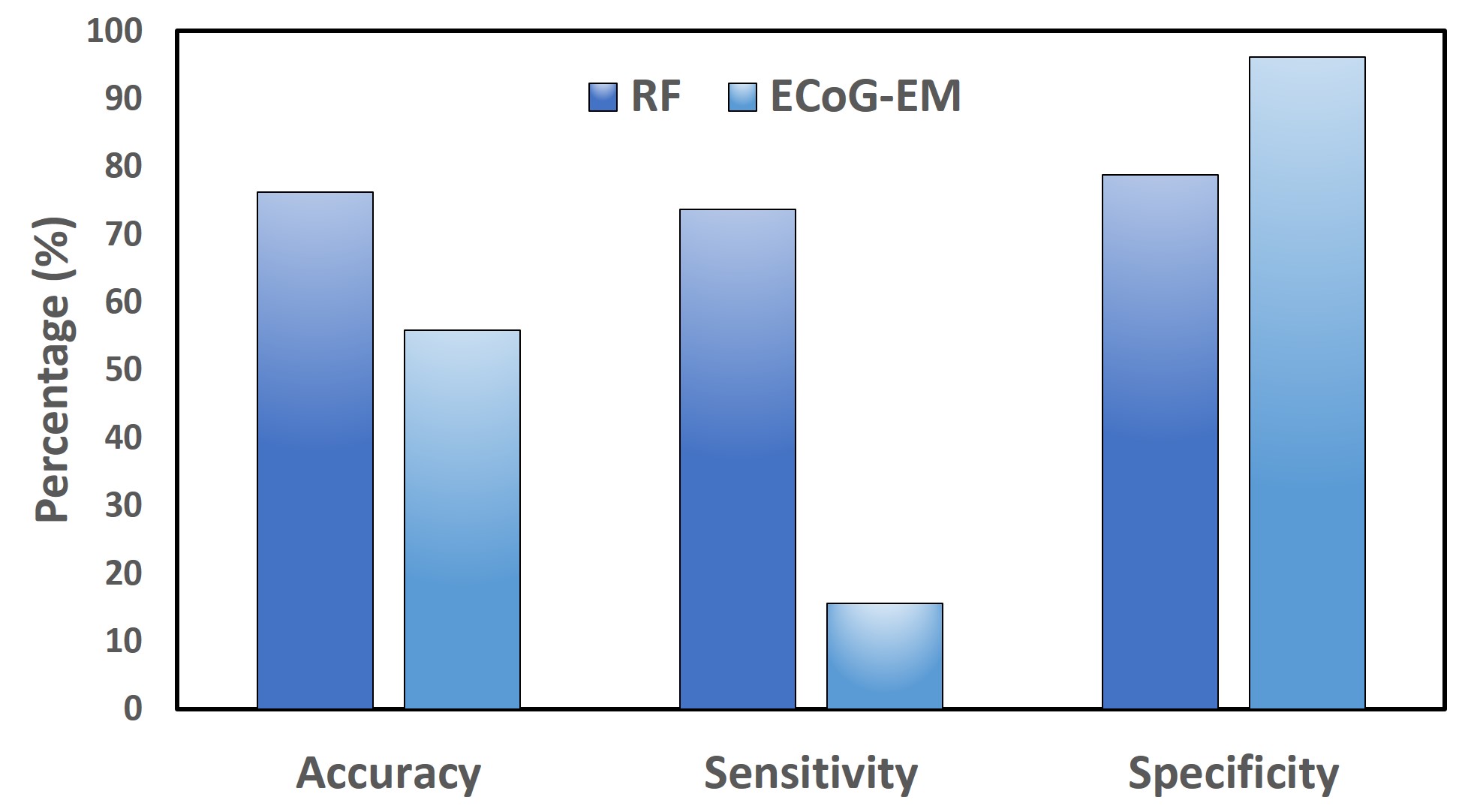}
\caption{Comparison of ECoG signal classification using proposed approach - Random Forest (RF) and conventionally used, ECoG-EM, on the language comprehension task.\label{sigfried}} 
\end{figure}

 ECoG-EM has been extensively tested on motor localization tasks \cite{prueckl2013cortiq}, but not as much on language localization. Still, ECoG-EM is considered to be the state of the art method. To have a fair comparison with ECoG-EM, we applied ECoG-EM on the frequently tested sub-bands - $\alpha,\beta$ and high-$\gamma$, as well as on the frequency bands beyond and upto 350 Hz. The results are shown in Figure~\ref{sigfried}. \textbf{While ECoG-EM approach provides a higher specificity, it has a much lower accuracy and sensitivity than the proposed RF based approach.} This is a strong validation of our hypothesis that discriminating PRCs and NRCs is a promising technique as compared to the baseline reference channel classification approach.  

\section{Conclusion}\label{conclusion}
Discriminating between the response in the eloquent language cortex regions based on the associated task is a challenging problem. In the current study, we developed a novel framework towards the ECoG-based eloquent cortex localization with promising results: 78\% accuracy on channel classification in comparison to the 55\% accuracy of the state of the art ECoG-based functional mapping. We showed the efficacy of machine learning based RTFM signal analysis as a strong alternative to the ESM. 

\ifCLASSOPTIONcompsoc
  \section*{Acknowledgments}
\else
  \section*{Acknowledgment}
\fi

The authors would like to acknowledge UCF-FH seed grant (PIs: M. Korostenskaja and U. Bagci) for supporting this  study. The authors would also like to thank Drs. Schott Holland and Jennifer Vannest for sharing the story listening task developed in their Neuroimaging Center at Cincinnati Children's Central Hospital. Special thanks to Drs. G. Schalk and P. Brunner for providing their in-house built version of BCI2000-based software for ECoG recording and for their continued support of our ECoG-related studies.




\end{document}